\documentclass[12pt,epsf]{article}
\pdfoutput=1
\usepackage[pdftex]{graphicx}
\usepackage{amsmath,amssymb}
\usepackage{bm}
\usepackage{ascmac}
\usepackage{braket}
\graphicspath{{./fig/}}
\usepackage{comment}
\usepackage{cite}

\newcommand{\beq}{\begin{equation}}
\newcommand{\eeq}{\end{equation}}
\newcommand{\bea}{\begin{eqnarray}}
\newcommand{\eea}{\end{eqnarray}}

\newcommand{\ba}{\begin{aligned}}
\newcommand{\ea}{\end{aligned}}

\def\pe2{p_E^2}

\begin{document}
\setlength{\baselineskip}{0.7cm}
\begin{titlepage}
\begin{flushright}
OCU-PHYS 539  \\
NITEP 102
\end{flushright}
\vspace*{10mm}%
\begin{center}{\LARGE\bf
Extranatural Flux Inflation
}
\end{center}
\vspace*{10mm}
\begin{center}
{\Large Takuya Hirose}$^{a}$ and 
{\Large Nobuhito Maru}$^{a,b}$, 
\end{center}
\vspace*{0.2cm}
\begin{center}
${}^{a}${\it
Department of Mathematics and Physics, Osaka City University, \\
Osaka 558-8585, Japan}
\\
${}^{b}${\it Nambu Yoichiro Institute of Theoretical and Experimental Physics (NITEP), \\
Osaka City University,
Osaka 558-8585, Japan}
\end{center}
\vspace*{1cm}

\begin{abstract}
We propose a new inflation scenario in flux compactification, 
 where a zero mode scalar field of extra components of the higher dimensional gauge field 
 is identified with an inflaton. 
The scalar field is a pseudo Nambu-Goldstone boson 
 of spontaneously broken translational symmetry in compactified spaces. 
The inflaton potential is non-local and finite, 
 which is protected against the higher dimensional non-derivative local operators 
 by quantum gravity corrections thanks to the gauge symmetry in higher dimensions 
 and the shift symmetry originated from the translation in extra spaces. 
We give an explicit inflation model in a six dimensional scalar QED, 
 which is shown to be consistent with Planck 2018 data. 
\end{abstract}
\end{titlepage}

\section{Introduction}

The hierarchy problem has been considered 
 as one of the guiding principles to study the physics beyond the Standard Model (SM) of particle physics.
In the SM, the quantum corrections to the mass of Higgs field are sensitive to 
 the square of the ultraviolet cutoff scale of the theory, 
 which is typically Planck scale or the scale of grand unified theory.
Since the cutoff scale is much larger than the 125 GeV Higgs mass, 
 solving the hierarchy problem requires an unnatural fine-tuning of parameters or 
 a new physics beyond the SM around TeV scale.
Although the latter approaches have been mainly studied so far, 
 no signature of new physics has been discovered, 
 which is likely to increase a discrepancy between the new physics scale and the Higgs mass. 
Therefore, it seems to be desirable such that the Higgs mass vanishes at the classical level 
 and is generated at the quantum effects by the dynamics different from the new physics one. 

As one of the approaches to the hierarchy problem, 
 the higher dimensional theory with magnetic flux compactification recently attracts attention.
Originally, the magnetic flux compactification has been studied in string theory \cite{BKLS, IU} 
 and many attractive properties has been known even in the field theory: 
 attempt to explain the number of the generations of the SM fermion \cite{Witten}, 
 computation of Yukawa coupling hierarchy \cite{CIM, 0903, MS}.
Remarkably, it has recently been shown that the quantum corrections to the masses of the scalar zero-mode field 
 originated from extra components of higher dimensional gauge field (called Wilson-line (WL) scalar field) are canceled 
 \cite{B1, Lee, B2, HM, 2-loop, HDO}. 
The physical reason of the cancellation is that the WL scalar field becomes a Nambu-Goldstone boson 
 of the spontaneously broken translational symmetry in compactified spaces 
 and the shift symmetry from the translation in compactified spaces forbids 
 the non-derivative terms as well as the mass term of the WL scalar field. 
Therefore, the massless nature of the WL scalar field is expected to be valid at any order of perturbation. 

In order to apply the scenario of the flux compactification 
 where the WL scalar field is identified with Higgs field to the hierarchy problem, 
 we need some mechanism to generate an explicit breaking term of the translational symmetry in compactified space 
 and the WL scalar field must be a pseudo NG boson such as pion.
In our previous paper \cite{finite}, 
 we have studied a possibility of realizing nonvanishing WL scalar mass in flux compactification.
By generalizing loop integrals of the quantum correction to WL scalar mass at one-loop,
 we derived the conditions for the loop integral and mode sum to be finite.
We further classified the four-point and three-point interaction terms 
 explicitly breaking the translational symmetry in compactified space 
 and generating the finite WL scalar mass at one-loop.
Using the simplest interactions of them, 
 we have explicitly shown the finite WL scalar mass at one-loop in a six dimensional scalar QED.  

Inflation is a very attractive scenario to solve many problems in the standard Big Bang cosmology 
 and its existence is supported by observations of cosmological parameters\cite{Planck}. 
Inflation has been considered to happen by a scalar field called as inflaton. 
Although many models of inflation has been proposed so far, 
 there is still no compelling model of inflation. 
In a slow-roll scenario of the inflation, 
 the scalar potential is required to be flat and stable under quantum gravity corrections, 
 which usually causes an unnatural fine-tuning of parameters of the theory 
 unless we have some dynamics or symmetry to control the inflaton dynamics. 
For instance, the inflaton in natural inflation \cite{natural} is identified with 
 the pseudo Nambu-Goldstone boson of some global symmetry.  
In extranatural inflation \cite{extranatural}, the inflaton is identified with the WL scalar field 
 of the gauge field in higher dimensions without magnetic flux. 
 In \cite{IKLM}, the inflaton and the curvaton are identified with the WL scalar fields 
 in a six dimensional gauge theory. 

In this paper, we propose a new scenario of inflation in flux compactification 
 where the WL scalar field is identified with an inflaton. 
Since this is a model of ``extranatural inflation" \cite{extranatural} with magnetic flux in compactified space, 
 we refer to our scenario as ``extranatural flux inflation".\footnote{The extranatural inflation is 
 an inflation model where the idea of gauge-Higgs unification is applied. 
 For the discussion of UV insensitivity for the WL scalar mass and potential, see \cite{GHU}} 
In constructing a slow-roll inflation model, 
 it is important to realize a flat inflaton potential stable 
 against quantum corrections and quantum gravity corrections 
 since the slow-roll conditions typically require the large field value larger than the Planck scale for the inflaton. 
Unless this stability is guaranteed by some physical reasons, 
 we cannot discuss the inflation by an effective field theory description 
 because it is beyond the range of their applicability. 
Similar to the extranatural inflation \cite{extranatural}, 
 the inflaton potential in our scenario is controlled by the gauge symmetry in higher dimensions. 
The non-derivative local operators of the WL scalar fields are forbidden by the gauge symmetry, 
 and by the shift symmetry originated from the translational symmetry in compactified spaces 
 in a context of flux compactification. 
Therefore, the higher dimensional operators suppressed by the Planck scale, 
 which is expected to be generated by the black holes, are forbidden. 
The quantum corrections to the local operators (by non-gravity interactions) are also forbidden. 
The remarkable nature is that the non-local Wlison line operators of the WL scalar field 
 are generated by quantum effects and become finite inspite of the non-renormalizable theory.   
This is a great advantage which is absent in other scenarios. 

As a simplest model, we consider a six dimensional scalar QED which was discussed in \cite{finite}. 
In this model, the WL scalar field is identified with the inflaton 
 and the one-loop Coleman-Weinberg potential for the WL scalar field is calculated. 
The potential is found to be finite as expected.  
Then, the slow-roll parameters, the spectral index, the scalar-to-tensor ratio, and the e-folding 
 are numerically computed and some viable parameter sets are found.

This paper is organized as follows. 
In the next section, our model is introduced and one-loop effective potential is calculated. 
Identifying the WL scalar field with an inflaton,  
 we discuss our inflation scenario by applying the one-loop effective potential to the inflaton potential 
 in section 3. 
In section 4, our numerical analysis is found 
 and some viable parameter sets consistent with Planck 2018 data are shown. 
A final section gives our summary. 

\section{Our setup}
In this section, we introduce our model and calculate the one-loop effective potential.

\subsection{Flux compactifiaction}
We consider a six-dimensional U(1) gauge theory with a constant magnetic flux couples to a scalar field in the bulk.
The six-dimensional spacetime is a product of four-dimensional Minkowski spacetime $M^4$ and two-dimensional torus $T^2$.
The Lagrangian we consider is
	\begin{align}
	\mathcal{L}=-\frac{1}{4}F_{MN}F^{MN}-(D_M\Phi)^*D^M\Phi
	+\kappa_6(\bar{\phi}\overline{\Phi}\Phi+\phi\overline{\Phi}\Phi),
	\label{ourlag}
	\end{align}
where the spacetime index is given by $M,N=0,1,2,3,5,6,~\mu,\nu=0,1,2,3,~m,n=5,6$ respectively 
 and we follow the metric convention as $\eta_{MN}=(-1,+1,\cdots,+1)$.
The field strength and the covariant derivative of U(1) gauge field $A_M$ are defined 
 by $F_{MN}=\partial_M A_N-\partial_N A_M,~D_M=\partial_M-ig_6 A_M$ 
 with a gauge coupling constant $g_6$ of mass dimension $-1$. 
$\Phi$ is a bulk scalar field.
$\phi$ is a scalar field related to the complex combination of $A_{5,6}$ and will be defined in detail later.
$\kappa_6$ is a dimensionless coupling constant.
The third term in \eqref{ourlag} is introduced to generates 
 nonvanishing quantum corrections to the mass of $\phi$ at one-loop \cite{finite}. 
In the context of this paper, 
 this term is also crucial to obtain a nonvanishing one-loop effective potential of $\phi$
 as well as the mass term. 
This is due to the property that the scalar field $\phi$ is the Nambu-Goldstone boson of 
 the translational symmetry in compactified spaces, 
 which forbids non-derivative terms such as the mass and potential terms 
 and then the explicit breaking terms for the translational symmetry are required. 

Let us introduce the magnetic flux in our model.
The magnetic flux is given by the nontrivial background (or vacuum expectation value (VEV)) 
 of fifth and sixth component of the gauge field $A_{5,6}$, 
 which must satisfy their classical equation of motion $\partial^m\braket{F_{mn}}=0$.
In our flux compactification, the background of $A_{5,6}$ is chosen as
	\begin{align}
	\braket{A_5}=-\frac{1}{2}fx_6,~~~\braket{A_5}=\frac{1}{2}fx_5,
	\end{align}
which introduces a constant magnetic flux density $\braket{F_{56}}=f$ with a real number $f$.
Note that this solution breaks an extra-dimensional translational invariance spontaneously.
Integrating over $T^2$, the magnetic flux is quantized as follows
	\begin{align}
	\frac{g_6}{2 \pi} \int_{T^{2}} d x_{5} d x_{6}\left\langle F_{56}\right\rangle=\frac{g_6}{2 \pi} L^{2} f=N \in \mathbb{Z},
	\end{align}
where $L^2$ is an area of two-dimensional torus.

It is useful to define $\partial,z,$ and $\phi$ as
	\begin{align}
	\partial\equiv \partial_z=\partial_5-i\partial_6,~~~z=\frac{1}{2}(x_5+ix_6),~~~\phi=\frac{1}{\sqrt{2}}(A_6+iA_5).
	\end{align}
Note that the VEV of $\phi$ is given by $\braket{\phi}=(\braket{A_6}+i\braket{A_5})/\sqrt{2}=f\bar{z}/\sqrt{2}$.
We expand $\phi$ around the flux background $\phi=\braket{\phi}+\varphi$, 
 where $\varphi$ is a quantum fluctuation.
To distinguish $\varphi$ from an introduced bulk scalar $\Phi$, 
 we call $\varphi$ Wilson line (WL) scalar field.

\subsection{Kaluza-Klein mass spectrum}
We need to derive Kaluza-Klein mass spectrum of the bulk scalar field $\Phi$ 
 for the calculation of effective potential.
To begin, we define the covariant derivatives in the complex coordinates $D,\bar{D}$ as
	\begin{align}
	D&=D_5-iD_6=\partial-\sqrt{2}g_6\phi=\mathcal{D}-\sqrt{2}g_6\varphi, \\
	\bar{D}&=D_5+iD_6=\bar{\partial}+\sqrt{2}g_6\bar{\phi}=\bar{\mathcal{D}}+\sqrt{2}g_6\bar{\varphi},\\
	\mathcal{D}&=\mathcal{D}_5-i\mathcal{D}_6=\partial-\sqrt{2}g_6\braket{\phi}, \\
	\bar{\mathcal{D}}&=\bar{\mathcal{D}}_5+i\bar{\mathcal{D}}_6=\bar{\partial}+\sqrt{2}g_6\braket{\bar{\phi}}.
	\end{align}
Next, we regard $\mathcal{D},\bar{\mathcal{D}}$ as creation and annihilation operators by
	\begin{align}
	a=\frac{1}{\sqrt{2g_6f}}i\bar{\mathcal{D}},~~~a^\dag=\frac{1}{\sqrt{2g_6f}}i\mathcal{D},
	\end{align}
which satisfy the commutation relation $[a,a^\dag]=1$.
This correspondence is an analogy to the quantum mechanics in magnetic field.
Hereafter, we denote $\alpha=2g_6f=4\pi N/L^2$.

We summarize the property of creation and annihilation operators.
The ground state mode functions $\xi_{0,j},~\bar{\xi}_{0,j}$ are determined by $a\xi_{0,j}=0,~a^\dag\bar{\xi}_{0,j}=0$, 
 where $j=0,\cdots,|N|-1$ accounts for the degeneracy of the ground state.
Creation and annihilation operators act on mode functions as
	\begin{align}
	a^\dag\xi_{n,j}=\sqrt{n+1}\xi_{n+1,j},~~~a\xi_{n,j}=\sqrt{n}\xi_{n-1,j}, \label{creatannihilate}
	\end{align}
and we can construct the higher mode function $\xi_{n,j}$ in the same way as the harmonic oscillator (in detail, see \cite{highermode})
	\begin{align}
	\xi_{n,j}=\frac{1}{\sqrt{n!}}(a^\dag)^n\xi_{0,j},~~~\bar{\xi}_{n,j}=\frac{1}{\sqrt{n!}}(a)^n\bar{\xi}_{0,j},
	\end{align}
where $n=0,1,2\cdots$ is Landau level.
The higher mode function satisfies an orthonormality condition
	\begin{align}
	\int_{T^2}d^2x \bar{\xi}_{n',j'}\xi_{n,j}=\delta_{n,n'}\delta_{j,j'}.
	\end{align}

Finally, we extract the mass term from the second term in \eqref{ourlag}.
Focusing on the extra-dimensional part of the second term in \eqref{ourlag}, we obtain
	\begin{align}
	\mathcal{L}_{scalar~mass}&=-\mathcal{D}_m\overline{\Phi} \mathcal{D}^m\Phi \nonumber \\
	&=-\overline{\Phi}\alpha\left(a^\dag a+\frac{1}{2}\right)\Phi.
	\end{align}
Noting that $a^\dag a$ is a number operator, the KK mass of bulk scalar field is given by
	\begin{align}
	m^2_{scalar}=\alpha\left(n+\frac{1}{2}\right). \label{scalarmass}
	\end{align}

\subsection{Four-dimensional effective Lagrangian}
Using the expansion of $\phi=\braket{\phi}+\varphi$, 
 the Lagrangian \eqref{ourlag} is accordingly deformed as
	\begin{align}
	\mathcal{L}&\supset-\frac{1}{4} F^{\mu \nu} F_{\mu \nu}
	-D_{\mu} \overline{\Phi} D^{\mu} \Phi-m^2_{scalar}\bar{\Phi}\Phi \nonumber \\
	&\quad- i g_6 \sqrt{2\alpha} \bar{\varphi} \bar{\Phi} a^{\dagger} \Phi+ i g_6 \sqrt{2\alpha} \varphi \bar{\Phi} a \Phi
	-2 g^{2}_6 \bar{\varphi} \varphi \overline{\Phi} \Phi \nonumber \\
	&\quad+\kappa_6(\bar{\varphi}\overline{\Phi}\Phi+\varphi\overline{\Phi}\Phi)+\kappa_6(\braket{\bar{\phi}}
	\overline{\Phi}\Phi+\braket{\phi}\overline{\Phi}\Phi),
	\end{align}
where we note that the unnecessary terms in our discussion are omitted.
To derive a four-dimensional effective Lagrangian by KK reduction, 
we need to expand $\Phi$ in terms of mode functions $\xi_{n,j}$
	\begin{align}
	\Phi=\sum_{n,j}\Phi_{n,j}\xi_{n,j}. \label{phiKKexpand}
	\end{align}
Integrating over $T^2$, the four-dimensional effective Lagrangian is obtained by
	\begin{align}
	\mathcal{L}_{4D}&=-\frac{1}{4} F^{\mu \nu} F_{\mu \nu}
	-\partial^{\mu} \bar{\varphi} \partial_{\mu} \varphi \nonumber \\
	&\quad+\sum_{n, j}\left(-D_{\mu} \overline{\Phi}_{n, j} D^{\mu} \Phi_{n, j}-\alpha\left(n+\frac{1}{2}\right)
	\overline{\Phi}_{n, j} \Phi_{n, j}\right. \nonumber \\
	&\quad- i g_4 \sqrt{2\alpha(n+1)} \overline{\varphi} \overline{\Phi}_{n+1, j} \Phi_{n, j}+ i g_4 \sqrt{2\alpha(n+1)} 
	\varphi \overline{\Phi}_{n, j} \Phi_{n+1, j}-2g^2_4\bar{\varphi}\varphi\overline{\Phi}_{n,j}\Phi_{n,j} \nonumber \\
	&\quad+\kappa_4\bar{\varphi}\overline{\Phi}_{n,j}\Phi_{n,j}
	+\kappa_4\varphi\overline{\Phi}_{n,j}\Phi_{n,j}+\kappa_4\braket{\phi}_I\overline{\Phi}_{n,j}\Phi_{n,j}
	+\kappa_4\braket{\bar{\phi}}_I\overline{\Phi}_{n,j}\Phi_{n,j}\Big), 
	\label{4Dlag}
	\end{align}
where $g_4=g_6/L$ is a four-dimensional gauge coupling constant 
and $\kappa_4=\kappa_6/L$ is a four-dimensional coupling constant. 
$\braket{\phi}_I$ and $\braket{\bar{\phi}}_I$ are expressed by
	\begin{align}
	\braket{\phi}_I=\int_{T^2}d^2 x \braket{\phi}\bar{\xi}_{n,j}\xi_{n',j'},~~~\braket{\bar{\phi}}_I
	=\int_{T^2}d^2 x \braket{\bar{\phi}}\bar{\xi}_{n,j}\xi_{n',j'}.
	\end{align}
When $\braket{\phi}=f\bar{z}/\sqrt{2}$, $\braket{\phi}_I$ and $\braket{\bar{\phi}}_I$ lead to zero 
 because of odd function with respect to integral variables $z$ or $\bar{z}$.
In the following, we omit the third and fourth terms in the fourth line of \eqref{4Dlag}.

\subsection{One-loop effective potential}
One-loop effective potential is described as
	\begin{align}
	V(\varphi,\bar{\varphi})=N \sum_{n=0}^\infty\int \frac{d^{4} k}{(2 \pi)^{4}}
	\ln\left(k^2+\alpha\left(n+\frac{1}{2}\right) + M^2(\varphi,\bar{\varphi})\right),
	\end{align}
where we have taken into account loop contributions from the bulk scalar field $\Phi$. 
$N$ is a number of the degeneracy.  
$M^2(\varphi,\bar{\varphi})$ is a field-dependent mass for the bulk scalar field $\Phi$. 

As for this $M^2(\varphi,\bar{\varphi})$, 
 we consider two limiting cases for a free parameter $U(1)$ gauge coupling, 
 namely $g_4 \ll 1$ and $g_4 \gg 1$. 
For that purpose, we read $M^2(\varphi,\bar{\varphi})$ from \eqref{4Dlag} as
	\begin{align}
	M^2(\varphi,\bar{\varphi})&=-\kappa_4\bar{\varphi}-\kappa_4\varphi + 2g^2_4\bar{\varphi}\varphi.
	\label{mass}
	\end{align}
While only the first two terms in (\ref{mass}) are considered in the $g_4 \ll 1$ case, 
 the last term proportional to $g^2_4$ in (\ref{mass}) is also considered in the $g_4 \gg 1$ case 
 in addition to the first two terms. 
In the case of $g_4 \simeq {\cal O}(1)$, 
 the terms linear in $g_4$ in (\ref{4Dlag}) should be also taken into account in $M^2(\varphi,\bar{\varphi})$. 
However, the obtained eigenvalues of $M^2(\varphi,\bar{\varphi})$ become complicated 
 and makes the computation of the effective potential hard.  
Therefore, we do not discuss this case in this paper.  
 

We can express the effective potential by using Schwinger's proper time as
	\begin{align}
	V&=- N \sum_{n=0}^\infty\int \frac{d^{4} k}{(2 \pi)^{4}}\int_{0}^{\infty}
	\frac{dt}{t}e^{-k^2 t-\alpha\left(n+\frac{1}{2}\right)t}e^{-M^2(\varphi,\bar{\varphi})t} \nonumber \\
	&=- N \frac{1}{16\pi^2}\int_{0}^{\infty}\frac{dt}{t^3}
	\frac{e^{-\frac{\alpha}{2}t}}{1-e^{-\alpha t}}e^{-M^2(\varphi,\bar{\varphi})t}. 
	\end{align}
To proceed a calculation of the effective potential further, 
 we notice an integral representation of Hurwitz zeta function 
	\begin{align}
	\zeta[s,a]=\frac{1}{\Gamma(s)}\int_{0}^{\infty}dt\frac{t^{s-1}e^{-at}}{1-e^{-t}},~~~\mathrm{Re}~s>1.  
	\label{Hurwitzintegral}
	\end{align}
Then, the effective potential and its derivatives by $\varphi$ can be expressed by
	\begin{align}
	V&=- N \frac{\alpha^2}{16\pi^2}\lim_{\epsilon\rightarrow0}\Gamma(\epsilon-2)\zeta
	\left[\epsilon-2,\frac{1}{2}+\frac{1}{\alpha}M^2(\varphi,\bar{\varphi})\right], \label{effectivept} \\
	V_{\varphi}&=- N \frac{\alpha\kappa}{16\pi^2}
	\lim_{\epsilon\rightarrow0} \Gamma(\epsilon-1) 
	\zeta\left[\epsilon-1,\frac{1}{2}+\frac{1}{\alpha}M^2(\varphi,\bar{\varphi})\right], \\
	V_{\varphi\bar{\varphi}}&= - N \frac{\kappa^2}{16\pi^2}
	\lim_{\epsilon\rightarrow0}\Gamma(\epsilon)\zeta\left[\epsilon,\frac{1}{2}+\frac{1}{\alpha}M^2(\varphi,\bar{\varphi})\right],
	\end{align}
where a parameter $\epsilon$ is introduced to regularize the integral of $t$. 
In particular, we can check that the $\epsilon\rightarrow0$ limit indeed agrees with 
 the results in case of $M^2(\varphi,\bar{\varphi})=0$ obtained in\cite{finite} 
 by diagrammatic calculations using the dimensional regularization.

\begin{figure}[http]
	\begin{center}
	\includegraphics[width=100mm]{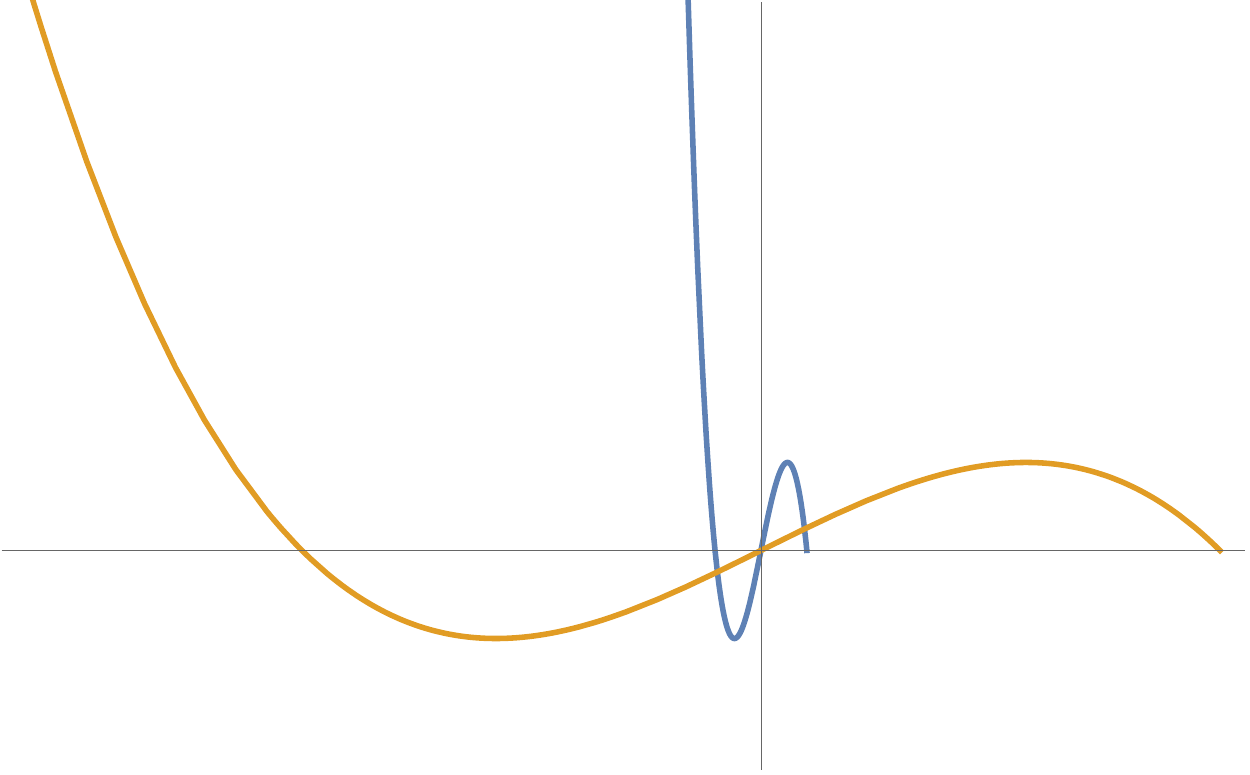}
	\end{center}
	\caption{Schematic picture of the effective potential in the case of $g_4\ll1$. 
	The blue and yellow lines shows $y=1.0\times10^{0}$, $y=1.0\times10^{-1}$ respectively.}
	\label{ptgraph}
	\end{figure}

In the $g_4\ll1$ case, we ignore $2g^2_4\bar{\varphi}\varphi$ in \eqref{mass} as mentioned above.
For convenience, we define the dimensionless variables in a four dimensional sense as
	\begin{align}
	z=\frac{\varphi}{M_P},~~~y=M_P\frac{\kappa_4}{\alpha}, 
	\end{align}
$M^2(\varphi,\bar{\varphi})/\alpha$ is then expressed by
	\begin{align}
	\frac{1}{\alpha}M^2(\varphi,\bar{\varphi})=-(z+\bar{z})y=-2xy, \quad \mathrm{Re}~z=x. 
       \label{g<1casemass}
	\end{align}
Thus, the effective potential is rewritten by
	\begin{align}
	V=- N \frac{\alpha^2}{16\pi^2}\lim_{\epsilon\rightarrow0}
	\Gamma(\epsilon-2)\zeta\left[\epsilon-2,\frac{1}{2}-2xy\right],
	\label{normalizedpt}
	\end{align}
and the effective potential is shown in figure \ref{ptgraph}.
If $L\sim M^{-1}_P$, the effective potential is close to flat as $y$ (or $\kappa_4$) takes smaller value.
Taking into account for the consistency with the original theory \cite{B1,B2}, 
 the small value of $y$ is favored.
 If $y\ll1$, $\kappa_4$ is small, which is independent of $g_4$.
 This implies that linear terms in $g_4$ can be neglected 
 because we can always take $g_4 \ll \kappa_4 L$. 

\begin{figure}[http]
	\begin{center}
	\includegraphics[width=100mm]{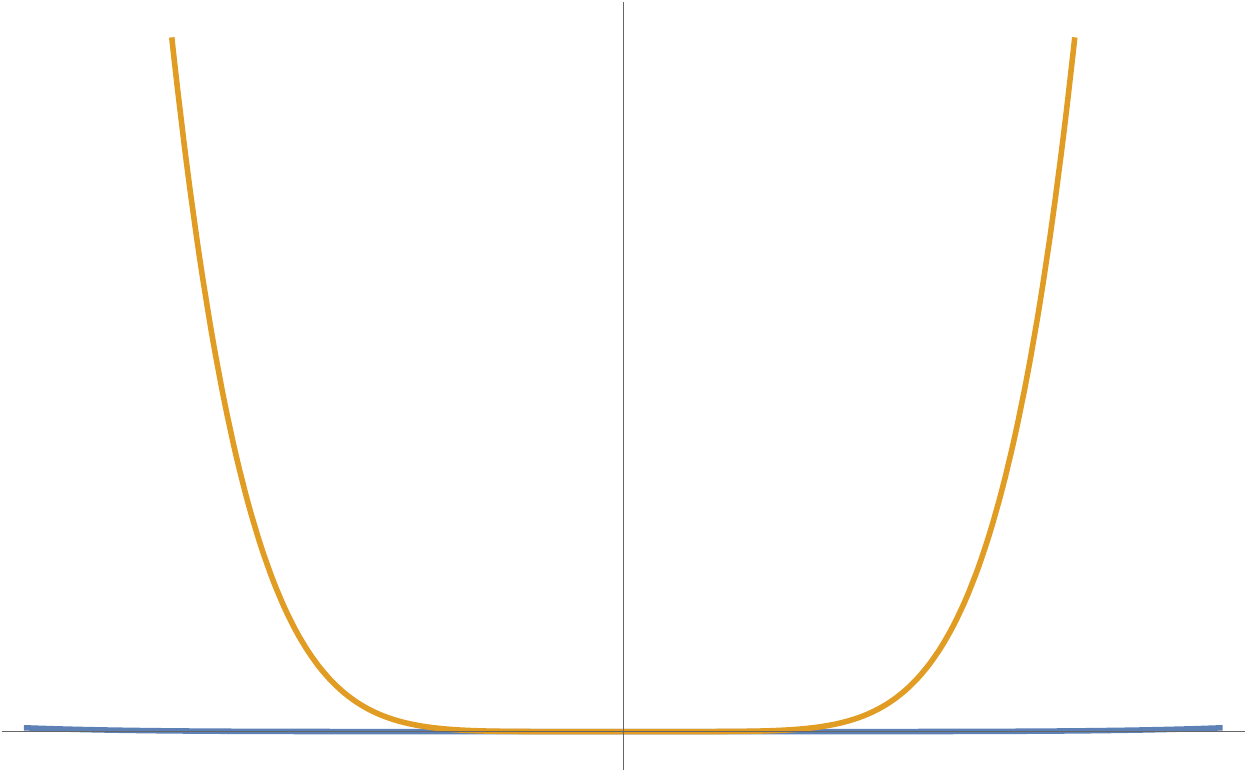}
	\end{center}
	\caption{Schematic picture of the effective potential in the case of $g_4\gg1$. 
	We take $y=1$ for simplicity.
	The yellow and blue lines shows $G=1.0\times10^{2}$, $G=1.0\times10^{3}$ respectively.}
	\label{ptgraph2}
	\end{figure}

In the $g_4\gg1$ case, $M^2(\varphi,\bar{\varphi})/\alpha$ is expressed by
	\begin{align}
	\frac{1}{\alpha} M^2(\varphi,\bar{\varphi})&=-(z+\bar{z})y+2\frac{g^2_4M^2_P}{\alpha}|z|^2 
	\nonumber \\
	&=-2uy+2G(u^2+v^2), 
	\label{g>1casemass}	
	\end{align}
 where $z\equiv u+i v$ and $G \equiv g^2_4 M^2_P/\alpha$ are defined in the second equality. 
Note that $G$ is almost an order of $g^2_4$ because $\alpha$ is independent of $g_4$.
Setting $u=v$ for simplicity, 
 the effective potential is given by
	\begin{align}
	V=- N \frac{\alpha^2}{16\pi^2}\lim_{\epsilon\rightarrow0}\Gamma(\epsilon-2)
	\zeta\left[\epsilon-2,\frac{1}{2}-2uy+4Gu^2\right],
	\end{align}
which is shown in figure \ref{ptgraph2}.
This effective potential in the case of $g_4\gg1$ behaves as $V\propto\Gamma[\epsilon-2]\zeta[\epsilon-2,4Gu^2]$.
Comparing with the potential in figure \ref{ptgraph}, 
 it seems difficult to apply the potential in figure \ref{ptgraph2} to an inflation model.

\section{Inflationary parameters}
Using the four-dimensional effective potential for the WL scalar field \eqref{effectivept}, 
 we propose a cosmological inflation model in flux compactifiaction, 
 where the WL scalar field is identified with an inflaton.

Slow-roll parameters $\epsilon$ and $\eta$ in our model are given by
	\begin{align}
	\epsilon &= \frac{M^2_P}{2}\left(\frac{V_\varphi}{V}\right)^2
	=\frac{M^2_P}{2}\left(\frac{\kappa_4}{\alpha}
	\lim_{\epsilon\rightarrow0}\frac{\Gamma(\epsilon-1)}{\Gamma(\epsilon-2)}
	\frac{\zeta\left[\epsilon-1,\frac{1}{2}+\frac{1}{\alpha}M^2(\varphi,\bar{\varphi})\right]}
	{\zeta\left[\epsilon-2,\frac{1}{2}+\frac{1}{\alpha}M^2(\varphi,\bar{\varphi})\right]}\right)^2, 
	\label{ep} \\
	\eta&=M^2_P\frac{V_{\varphi\varphi}}{V}
	=M^2_P\left(\frac{\kappa_4^2}{\alpha^2}\lim_{\epsilon\rightarrow0}
	\frac{\Gamma(\epsilon)}{\Gamma(\epsilon-2)}
	\frac{\zeta\left[\epsilon,\frac{1}{2}+\frac{1}{\alpha}M^2(\varphi,\bar{\varphi})\right]}
	{\zeta\left[\epsilon-2,\frac{1}{2}+\frac{1}{\alpha}M^2(\varphi,\bar{\varphi})\right]}\right).  
	\label{eta}
	\end{align}
Noting that Hurwitz zeta function can be expressed by Bernoulli polynomials $B_n(x)$ as follows
	\begin{align}
	\zeta[-n,x]=-\frac{B_{n+1}(x)}{n+1},
	\end{align}
we can further simplify \eqref{ep} and \eqref{eta},  
	\begin{align}
	\epsilon&=\frac{y^2}{2}
	\left(-2\frac{\zeta\left[-1,\frac{1}{2}+\frac{1}{\alpha}M^2(\varphi,\bar{\varphi})\right]}
	{\zeta\left[-2,\frac{1}{2}+\frac{1}{\alpha}M^2(\varphi,\bar{\varphi})\right]}\right)^2
	=\frac{9y^2}{2}\left(\frac{B_2(\frac{1}{2}+\frac{1}{\alpha}M^2(\varphi,\bar{\varphi}))}
	{B_3(\frac{1}{2}+\frac{1}{\alpha}M^2(\varphi,\bar{\varphi}))}\right)^2, \\
	\eta&=y^2\left((-1)(-2)
	\frac{\zeta\left[0,\frac{1}{2}+\frac{1}{\alpha}M^2(\varphi,\bar{\varphi})\right]}
	{\zeta\left[-2,\frac{1}{2}+\frac{1}{\alpha}M^2(\varphi,\bar{\varphi})\right]}\right) 
	=6y^2\frac{B_1(\frac{1}{2}
	+\frac{1}{\alpha}M^2(\varphi,\bar{\varphi}))}{B_3(\frac{1}{2}+\frac{1}{\alpha}M^2(\varphi,\bar{\varphi}))}. 
	\end{align}
Slow-roll conditions to realize inflation require $\epsilon \ll 1, |\eta| \ll 1$.  

The number of e-folding before the end of inflation is
	\begin{align}
	N_*=\frac{1}{M^2_P}\int_{\varphi_{f}}^{\varphi_*}\frac{V}{V_{\varphi}}d\varphi
	=\frac{2}{3y}\int_{\varphi_*}^{\varphi_f}
	\frac{B_3\left(\frac{1}{2}+\frac{1}{\alpha}M^2(\varphi,\bar{\varphi})\right)}
	{B_2\left(\frac{1}{2}+\frac{1}{\alpha}M^2(\varphi,\bar{\varphi})\right)}  d\varphi.
	\end{align}
To solve the horizon and flatness problems, the number of e-folding $N_*$ has to be at least $50<N_*<60$.
$\varphi_f$ is the value of the end of inflation determined by $\epsilon(\varphi_f)=1$, 
 which violates the slow-roll conditions.
$\varphi_*$ is determined so that the e-folding can satisfy $50<N_*<60$.

The spectral index and the tensor-to-scalar ratio are given in a slow-roll approximation as
	\begin{align}
	n_s=1-6\epsilon+2\eta,~~~r=16\epsilon.
	\end{align}
Planck 2018 data \cite{Planck} gives constraints on $n_s=0.9649 \pm 0.0042$ and $r < 0.10$.

\section{Numerical results}
In this section, our numerical results are shown.

\subsection{$g_4\ll$1 case}
In this case, $M^2(\varphi,\bar{\varphi})/\alpha$ corresponds to \eqref{g<1casemass}, 
 where the slow-roll parameters $\epsilon$ and $\eta$ are provided by
	\begin{align}
	\epsilon&=\frac{9y^2}{2}\left(\frac{B_2(\frac{1}{2}-2xy)}{B_3(\frac{1}{2}-2xy)}\right)^2,~~~
	\eta=6y^2\frac{B_1(\frac{1}{2}-2xy)}{B_3(\frac{1}{2}-2xy)}.
	\end{align}
To compute the e-folding $N_*$, we need to know the value of end of inflation $x_f=\varphi_f/M_P$, 
 which is determined by the condition of the end of inflation $\epsilon(x_f)=1$.
The number of e-folding is 
	\begin{align}
	N_*=\frac{2}{3y}\int_{x_i}^{x_f}\frac{B_3(\frac{1}{2}-2xy)}{B_2(\frac{1}{2}-2xy)}dx,
	\end{align}
where $x_i=\mathrm{Re}\varphi_*/M_P$.
Sample of our numerical solutions $x_i, x_f, N_*$ at some points of $y$ are shown in Table \ref{table1}, 
 where the e-folding $N_*=50, 60$ are taken. 
One might think that our results are not reliable 
 since the WL scalar field value is quite larger than the Planck scale, 
 which is beyond an applicability of the effective field theory. 
As mentioned in introduction, 
 the gauge symmetry in our theory is not broken by quantum gravity effects 
 and forbids any dangerous higher dimensional local operators suppressed by the Planck scale 
 as well as the non-derivative local operators of the WL scalar field. 
Therefore, our results are reliable. 
\begin{table}[htb]
\begin{center}
\begin{tabular}{c|c|c|c|c}
& $y$ & $x_i$ & $x_f$ & $N_*$ \\ \hline
$A_{50}$ & $1.0\times10^{-2}$ & $\quad-30.4669\quad$ & $
\quad-25.3611\quad$ & $\quad50.002\quad$ \\ \hline
$A_{60}$ & $1.0\times10^{-2}$ & $-31.0285$ & $-25.3611$ & 60.0098 \\ \hline
$B_{50}$ & $5.0\times10^{-3}$ & $-55.2516$ & $-50.3573$ & 50.0016 \\ \hline
$B_{60}$ & $5.0\times10^{-3}$ & $-55.7738$ & $-50.3573$ & 60.0004  \\ \hline
$C_{50}$ & $1.0\times10^{-3}$ & $-255.063$ & $-250.354$ & 50.0144 \\ \hline
$C_{60}$ & $1.0\times10^{-3}$ & $-255.549$ & $-250.354$ & 60.0162  \\ \hline
$D_{50}$ & $5.0\times10^{-4}$ & $-505.038$ & $-500.354$ & 50.0099  \\ \hline
$D_{60}$ & $5.0\times10^{-4}$ & $-505.519$ & $-500.354$ & 60.0087 \\ \hline
$E_{50}$ & $1.0\times10^{-4}$ & $-2505.018$ & $-2500.35$ & 50.0098 \\ \hline
$E_{60}$ & $1.0\times10^{-4}$ & $-2505.495$ & $-2500.35$ & 60.0078 \\
\end{tabular}
\caption{Sample of our numerical solutions $x_i, x_f, N_*$ at some points of $y$.}
\label{table1}
\end{center}
\end{table}
%

Using the numerical solutions in Table \ref{table1}, 
 the slow-roll parameters $\epsilon, \eta$, the spectral index $n_s$, 
 and the scalar-to-tensor ratio $r$ are calculated and shown in Table \ref{table2}.  
Comparing our results in Table \ref{table2} with $n_s$ and $r$ in Planck 2018 data, 
 our results are found to be relatively good agreement with the data. 
If $y$ is taken to be a large value such as $y=1.0\times10^{2}$, 
 $n_s$ and $r$ cannot be satisified with Planck 2018 data.
\begin{table}[htb]
\begin{center}
\begin{tabular}{c|c|c|c|c} 
& $\epsilon$ & $\eta$ & $n_s$ & $r$ \\ \hline
$A_{50}$ & $\quad0.00683107\quad$ & $\quad0.00494671\quad$ & $\quad0.968907\quad$ & $\quad0.109297\quad$ \\ \hline
$A_{60}$ & 0.00582958 & 0.00444092 & 0.973904 & 0.0932733  \\ \hline
$B_{50}$ & 0.00594149 & 0.00271376 & 0.969779 & 0.0950639 \\ \hline
$B_{60}$ & 0.00502904 & 0.00245613 & 0.974738 & 0.0804646 \\ \hline
$C_{50}$ & 0.00517205 & 0.000586594 & 0.970141 & 0.0827528 \\ \hline
$C_{60}$ & 0.00432933 & 0.000534704 & 0.975093 & 0.0692693  \\ \hline
$D_{50}$ & 0.00507359 & 0.000296245 & 0.970151 & 0.0811775 \\ \hline
$D_{60}$ & 0.00423959 & 0.000270297 & 0.975103 & 0.0678334  \\ \hline
$E_{50}$ & 0.00499408 & 0.0000597248 & 0.970155 & 0.0799053 \\ \hline
$E_{60}$ & 0.00416705 & 0.0000545352 & 0.975107 & 0.0666728  \\
\end{tabular}
\caption{Inflation parameters $\epsilon, \eta, n_s, r$ obtained from our model.}
\label{table2}
\end{center}
\end{table}
%
%

	\begin{figure}[http]
	\begin{center}
	\includegraphics[width=90mm]{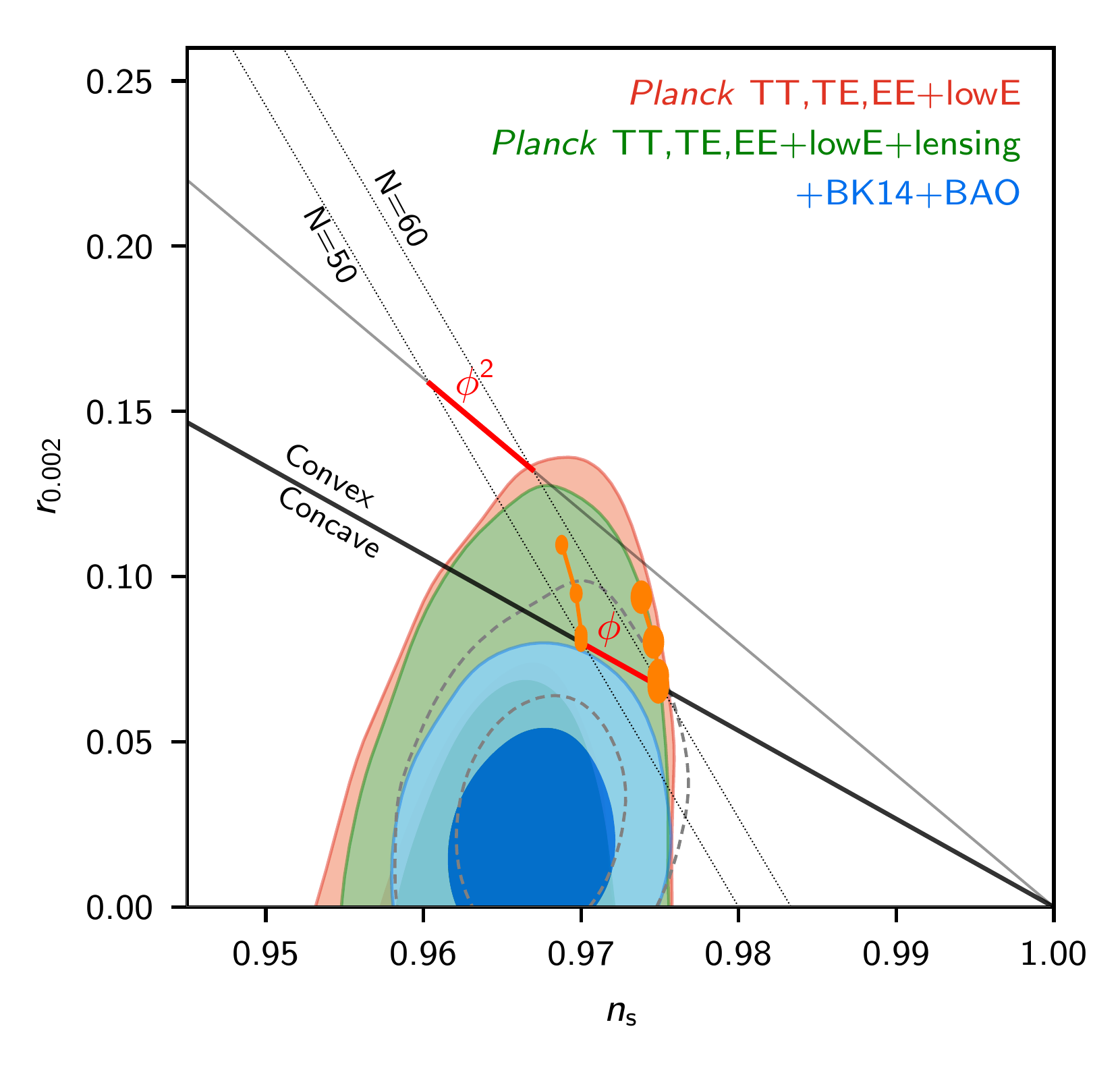}
	\end{center}
	\caption{Our results (table \ref{table2}) in the $n_s$-$r$ plot from Planck 2018 data \cite{Planck}.
	Orange circles are our results, 
	where the small and large ones represent $N_*=50$ and $N_*=60$, respectively.}
	\label{result}
	\end{figure}
Our results are shown in $(n_s, r)$ plot of Figure \ref{result} from Planck 2018 data \cite{Planck}.
Orange circles are our results where small and large ones corresponds to $N_*=50$ and $N_*=60$, respectively.
As the parameter $y$ is decreased, our results in $(n_s, r)$ plot go downward. 
Our results are within a parameter region indicating the combining data of Planck TT, TE, EE+lowE+lensing at CL95\%.

From the parameter $y$, we can estimate the value of $\kappa_6$, 
 which provides the compactification scale and the 6D Planck scale $M_{6P}$ in our model. 
\begin{table}[htb]
\begin{center}
\begin{tabular}{c|c|c}
& $L$ [GeV$^{-1}$]$(\kappa_6=10^x L)$ & $M_{6P}=\sqrt{M_P/L}$\\
\hline  
$y=1.0\times10^{-2}$ & $\quad3.20941\times10^{-x/2-10}\quad$ & $\quad1.9497\times10^{x/4+14}\quad$ \\ \hline
$y=5.0\times10^{-3}$ & $2.26939\times10^{-x/2-10}$ & $2.3186\times10^{x/4+14}$ \\ \hline
$y=1.0\times10^{-3}$ & $1.0149\times10^{-x/2-10}$ & $3.46711\times10^{x/4+14}$ \\ \hline
$y=5.0\times10^{-4}$ & $7.17646\times10^{-x/2-11}$ & $4.12311\times10^{x/4+14}$ \\ \hline
$y=1.0\times10^{-4}$ & $3.20941\times10^{-x/2-11}$ & $6.16549\times10^{x/4+14}$ \\
\end{tabular}
\caption{The value of $L,M_{6P}$}
\label{addtable}
\end{center}
\end{table}
$\kappa_6 L$ is determined by $y$ as follows. 
	\begin{align}
	&y=M_P\frac{\kappa_4}{\alpha}=M_P\frac{\kappa_6/L}{4\pi N/L^2}=\frac{M_P L\kappa_6}{4\pi N} \nonumber \\
	\Leftrightarrow& \kappa_6 =4\pi N\frac{y}{M_PL}
	\end{align}
where the number of degeneracy is assumed to be $N=10$. 
If we assume $\kappa_6=m_{inflaton}L$ and $m_{inflaton}=10^x$ GeV, $L$ is estimated.
$\kappa_6 L,~L,~M_{6P}$ are shown in Table \ref{addtable}.

Now, we discuss how small the gauge coupling $g_4$ is required for a successful inflation. 
Although the gauge coupling itself $g_4$ is a free parameter, 
 the constraint from slow-roll parameter condition can be obtained through the coupling constant $\kappa_6$,
which can be derived from $\epsilon\ll1$, 
	\begin{align*}
	&\frac{3}{\sqrt{2}}yB_2(1/2-2x_i y)\ll B_3(1/2-2x_i y), 
	\end{align*}
which implies,
	\begin{align}
	y\ll\sqrt{\frac{2\sqrt{2}x_i+1}{4(12x_i^2+8\sqrt{2}x_i^3)}}. 
	\end{align}
In the condition $\epsilon\ll1$, 
$y \ll 1$ is immediately found. 
Therefore, we obtain $\kappa_6\ll10^{-19}/L$, 
 which means $\kappa_6 \ll 1$ because the maximum value of $10^{-19}/L$ is at $1/L\sim M_P$.
As mentioned in section 2.4, we can always take the free parameter $g_4$ less than $\kappa_6$.
For a successful inflation in our model, 
 we have only to take the free parameter gauge coupling $g_4$ with $g_4 \ll \kappa_6$ 
 and this can be always possible.

\subsection{$g_4\gg$1 case}
In this case, $M^2(\varphi,\bar{\varphi})$ corresponds to \eqref{g>1casemass}.
Under $u=v$, $\epsilon$ and $\eta$ are expressed by
	\begin{align}
	\epsilon=\frac{9y^2}{2}\left(\frac{B_2(\frac{1}{2}+4Gu^2)}{B_3(\frac{1}{2}+4Gu^2)}\right)^2,~~~
	\eta=6y^2\frac{B_1(\frac{1}{2}+4Gu^2)}{B_3(\frac{1}{2}+4Gu^2)},
	\end{align}
where we ignore $-2uy$ because $y$ is small.
The number of e-folding is
	\begin{align}
	N_*=\frac{2}{3y}\int_{u_i}^{u_f}\frac{B_3(\frac{1}{2}+4Gu^2)}{B_2(\frac{1}{2}+4Gu^2)}du.
	\end{align}

As in the $g_4\ll1$ case, We obtain the value of $u_i$ and $u_f$ for a value of $G$.
Taking $G=1,0\times10^{3}$ as an example, we find $u_i=-3.8403$ and $u_f=-0.7282$.
Using these values, $n_s$ and $r$ are $n_s=0.99569$ and $r=0.0206896$.
$r$ is consistent with Planck 2018 constraint, but $n_s$ is not.
Thus, comparing with the potential in $g_4\ll1$ case, 
the potential in the $g_4\gg1$ case is not suitable for the inflation.

\subsection{The vacuum energy during inflation}
In order for our model to be consistent with inflationary setup, 
the vacuum energy during inflation should be smaller than 4D Planck scale and the compactification scale. 
We verify this requirement. 
As you can see Table \ref{table1}, $x_fy$ takes 1/4 during inflation.
Thus, the vacuum energy becomes
	\begin{align}
	V_{vac}&=\braket{V}=-N\frac{\alpha^2}{16\pi^2}\lim_{\epsilon\rightarrow0}\Gamma(\epsilon-2)\zeta[\epsilon-2,1/2] \nonumber \\
	&=-N\frac{\alpha^2}{16\pi^2}\left(\frac{1}{2}\zeta^{(1,0)}[-2,1/2]\right) \nonumber \\
	&=-\frac{3N^3\zeta(3)}{32\pi^2}\frac{1}{L^4}
	\end{align}
where we take into account in the second equality that the VEV of inflaton field is zero during inflation. 
Setting $N=10$, $V_{vac}$ is estimated to be 
$\mathcal{O}(10)\times L^{-4} \sim \mathcal{O}(10) \times \left( \frac{M_{6P}}{M_P} \right)^4 M_{6P}^4$. 
In large extra dimensions, 6D Planck scale is smallar than 4D Planck scale $M_{6P}<M_P$, 
unless the compactification scale is the 4D Planck scale. 
Therefore, $1/L^4 < | V_{vac} | < M^4_P$ are satisfied as long as the compactification scale is smaller than 4D Planck scale.

\section{Summary}
In this paper, we have proposed a new inflation model in flux compactification, 
 which is refered to as ``extranatural flux inflation". 
In this model, the WL scalar field, 
 which is originated from extra components of the gauge field in higher dimensions, 
 is identified with an inflaton field. 
The great advantage of our model is that the inflation potential is protected 
 by the gauge symmetry of the theory 
 from the dangerous higher dimensional local operators by quantum gravity effects. 
Our inflation potential is nonlocal and finite, which is generated by quantum corrections. 
Therefore, the extranatural inflation with flux is very predictable regardless of the non-renormalizable theory. 
We have considered a model of six dimensional scalar QED in flux compactification
 and calculated an inflation potential. 
We have shown that our model in the weak gauge coupling case  
 is consistent with Planck 2018 data.

\section*{Acknowledgments}
We would like to thank Kazumasa Okabayashi for useful discussions and comments.


\end{document}